\documentclass[aip,rsi,amsmath,amssymb,reprint]{revtex4-1}
\usepackage{graphicx,color}  
\usepackage{dcolumn}
\usepackage{bm}

\begin{document}

\preprint{AIP/123-QED}

\title[Simulation study of bacterial chromosome compression]{Entropic elasticity and dynamics of the bacterial chromosome:\\ a simulation study}

\author{M.~C.~F. Pereira}
\affiliation{SUPA, School of Physics and Astronomy, University of Edinburgh, Peter Guthrie Tait Road, Edinburgh EH9 3FD, Edinburgh, UK}

\author{C.~A. Brackley}
\affiliation{SUPA, School of Physics and Astronomy, University of Edinburgh, Peter Guthrie Tait Road, Edinburgh EH9 3FD, Edinburgh, UK}

\author{J.~S. Lintuvuori}
\affiliation{Laboratoire de Physique des Solides, CNRS, Univ. Paris-Sud, Universite Paris-Saclay, 91405 Orsay Cedex, France}

\author{D. Marenduzzo}
\affiliation{SUPA, School of Physics and Astronomy, University of Edinburgh, Peter Guthrie Tait Road, Edinburgh EH9 3FD, Edinburgh, UK}

\author{E. Orlandini}
\affiliation{Dipartimento di Fisica e Astronomia and Sezione INFN, Universit\`a di Padova, via Marzolo 8, Padova, 35131 PD, Italy}

\date{\today}

\begin{abstract}
We study the compression and extension dynamics of a DNA-like polymer interacting with non-DNA binding and DNA-binding proteins, by means of computer simulations. The geometry we consider is inspired by recent experiments probing the compressional elasticity of the bacterial nucleoid (DNA plus associated proteins), where DNA is confined into a cylindrical container and subjected to the action of a ``piston'' -- a spherical bead to which an external force is applied. We quantify the effect of steric interactions (excluded volume) on the force-extension curves as the polymer is compressed. We find that non-DNA-binding proteins, even at low densities, exert an osmotic force which can be a lot larger than the entropic force exerted by the compressed DNA. The trends we observe are qualitatively robust with respect to changes in protein size, and are similar for neutral and charged proteins (and DNA). We also quantify the dynamics of DNA expansion following removal of the ``piston'': while the expansion is well fitted by power laws, the apparent exponent depends on protein concentration, and protein-DNA interaction in a significant way. We further highlight an interesting kinetic process which we observe during the expansion of DNA interacting with DNA-binding proteins when the interaction strength is intermediate: the proteins bind while the DNA is packaged by the compression force, but they ``pop-off'' one-by-one as the force is removed, leading to a slow unzipping kinetics. Finally, we quantify the importance of supercoiling, which is an important feature of bacterial DNA {\it in vivo}.
\end{abstract}

\maketitle

\section{\label{sec:introduction}Introduction}

The genome of living organisms, from bacteria to humans, is under remarkable confinement in physiological conditions~\cite{alberts,confinedDNA}. For instance, the circular chromosome of {\it E. coli} would be over 1 mm long if stretched out, yet it needs to fit within the bacterial cell which is a $2\times 1 \times1$ $\mu$m ellipsoid. Likewise, there is about $2$ m of DNA in a single human nucleus, whose typical size is only about 10 $\mu$m. Another useful way to quantify the degree of confinement of the bacterial DNA is to estimate its gyration radius, $R_g$, at equilibrium, which is about 5 $\mu$m; since this is larger than the bacterial cell, the bacterial genome is in the ``semi-dilute'' regime of polymer physics~\cite{DoiEdwards}. 

There are at least four mechanisms through which the bacterial chromosome is compacted within the cell~\cite{alberts,Joyeux2015}. First, and most obviously, it is confined within the cell wall; note however that the chromosome does not occupy the entire cell so it must be compacted further. Second, there is a depletion attraction between genome segments induced by the crowding of non-DNA-binding macromolecules~\cite{deVries2012}. Third, the genome is associated with a number of architectural or nucleoid associated proteins (NAPs)~\cite{Browning2010,Song2015} which can bind the DNA at more than one point, creating effective DNA-DNA attractive interactions which help to reduce the space occupied by the chromosome. Fourth, the bacterial chromosome is a supercoiled loop: i.e., the helical pitch is different from the one favoured thermodynamically -- 10.5 base pairs (bps) for B-DNA~\cite{alberts}. In practice, bacterial DNA is negatively supercoiled~\cite{DNAtopology}, so the helix is slightly underwound. The degree of negatively supercoiling is about 5$\%$, which means that in a length of DNA, which would have 20 turns at thermodynamic equilibrium, there are only 19 turns. The twist deficit can be converted into negative writhe, which creates a local folding of the DNA, again favouring compaction.

As a first approximation, one may consider only the first of these compaction mechanisms and view the bacterial chromosome as a biopolymer under tight confinement~\cite{JunReview}. According to this model, there is a large decrease in entropy when the DNA is within the cell, and one expects this to create an entropic pressure, or force, on the confining walls~\cite{Pelletier,JunMulder,Jun2008,Jung2012}. This naturally explains why the bacterial chromosome tends to expand when the confining cell wall is removed. The entropic force exerted by the DNA was measured in an interesting experiment by Pelletier {\it et al.}~\cite{Pelletier}. In that work, single bacterial cells were trapped in an array of cylindrical cavities, with diameter just larger than the width of the bacteria, and the height much larger than its length. After the cells were trapped, their walls were lysed so that the enclosed DNA was free to expand and to increase its conformational entropy. The DNA was found to reach a height about ten times larger than that of the bacterium. The entropic force exerted by the expanding chromosome on a colloidal bead placed in the channel on top of the DNA was measured by means of optical tweezers, and it was found that this force was much larger than expected on the basis of a simple theory (reviewed in Section~\ref{sec:theory}) for Gaussian -- i.e., infinitely thin -- polymers, with the known persistence length of DNA, 50 nm. 

In this work, we present a simulation study of a situation closely related to the experiment discussed above~\cite{Pelletier}. With respect to previous theoretical work on the bacterial chromosome based on polymer physics~\cite{JunMulder,Jun2008,Junier2013,Mirny,LagomarsinoReview,Scolari,Jung2012,Jung2013,Kim2015,Shendruk2014,Wegner2016}, the main novelty here is that we include in our simulations the effects of bacterial proteins, whether binding to DNA or not, and we quantify their effect on the compression elasticity of DNA and its dynamics.  

More specifically, we analyse the entropic compression elasticity and expansion dynamics of a model bacterial DNA confined in a thin cylinder and subject to the action of a ``piston'', a colloidal bead under the influence of an external force. This set-up allows us to measure the force experienced by the piston as a function of DNA compression. We find that steric effects and the presence of proteins strongly affect both the entropic elasticity of the DNA under compression, and also the dynamics of its extension once the piston is removed. In particular, non-DNA-binding proteins exert an osmotic pressure which can be at least as large as the entropic force exerted by the DNA; the macromolecular crowding they introduce is also important in determining the polymer dynamics. DNA-binding proteins can further lead to interesting ``popping-off'' kinetics, when the thermodynamic interactions with the DNA is tuned such that it leads to stable binding in confinement, but not in solution. Finally, we find that supercoiling and DNA topology have a relatively little effect on force-extension curves, but can affect the compression dynamics significantly.

Our results complement the simulations of Refs.~[\onlinecite{Kim2015,Shendruk2014}] which study the compaction of a confined bacterial DNA due to non-DNA-binding proteins, and quantify how the polymer collapse depends on crowder size, concentration, and polydispersity -- here we additionally quantify how crowding affects the out-of-equilibrium expansion dynamics. Other relevant, yet distinct, works are those of Refs.~[\onlinecite{Joyeux2015,deVries2012,Wegner2016}] which quantified the extent to which H-NS and other DNA-binding proteins compactifies DNA. 

\section{\label{sec:MaterialsandMethods}Materials and Methods}

\subsection{\label{sec:model}Model: set-up and potentials used}

The system we consider consists of a bacterial DNA molecule confined in a cylindrical pore, that is compressed by a piston in the absence or presence of proteins. DNA is modelled as a linear (Sections~\ref{sec:sim-compressDNA}-\ref{sec:sim-popoff}) or circular (Section~\ref{sec:sim-supercoiling}) self-avoiding polymer composed of spherical beads with diameter $\sigma$. Our coarse grained model and force fields are similar to those used to study bacterial DNA, or DNA-protein systems in Refs.~[\onlinecite{JunMulder,Kim2015,Brackley2013,LeTreut2016}].

The interaction potential between monomers is defined by three contributions. First, two neighbouring monomers are bound via a finitely extensible non-linear elastic (FENE) spring given by the potential 
\begin{eqnarray}
U_{{\rm FENE}}(r_{i,i+1})=-\frac{K_{\rm FENE}r_0^2}{2}\ln \left[ 1 - \left( \frac{r_{i,i+1}}{r_0} \right)^2 \right],
\end{eqnarray}
where $r_{i,i+1}$ is the distance between the {\it i}th bead and its nearest neighbour (the ({\it i+1})th) along the chain, $r_0=1.6\sigma$ is the maximal extent of the bond and $K_{{\rm FENE}}=30k_BT/\sigma^2$ is the bond energy. Second, there is a steric (excluded volume) interaction between all beads that is set by the Weeks-Chandler-Andersen potential, as follows:
\begin{eqnarray}\label{V_WCA}
U_{{\rm WCA}}(r_{ij})= 4k_{B}T \left[ \left( \frac{d_{ij}}{r_{ij}} \right)^{12} - \left( \frac{d_{ij}}{r_{ij}} \right)^{6} \right] +  k_{B}T
\end{eqnarray}
for $r_{ij}<2^{1/6}d_{ij}$, and $U_{{\rm WCA}}(r_{ij})=0$ otherwise. Here $k_BT$ is the thermal energy ($k_B$ is the Boltzmann constant and $T$ is the temperature), $r_{ij}$ the distance between the {\it i}th and {\it j}th beads and $d_{ij}$ the mean of the diameters of the two interacting beads, i.e., $d_{ij}=\sigma$. [Note that under this choice of FENE and WCA potentials, the DNA bond length is approximately equal to $\sigma$.] Third, the bending rigidity of the polymer is introduced by a Kratky-Porod potential for every three adjacent monomers 
\begin{eqnarray}\label{KratkyPorod}
U_{{\rm BEND}}(\theta)=K_{{\rm BEND}}(1+\cos (\theta)),
\end{eqnarray}
where $\theta$ is the angle between the three consecutive monomers and $K_{{\rm BEND}}$ the bending energy. $K_{{\rm BEND}}$ sets the flexibility of the polymer, since it determines the persistence length $l_p$ (in units of $\sigma$): $l_p=K_{\rm BEND}/k_BT$. We use the well characterised persistence length for naked double-stranded DNA $l_p=20\sigma=50$ nm.

In the simulations where supercoiling is included, a circular polymer has to be considered, and three more interaction potentials taken into account. DNA supercoiling results from over- or under-twisting the helical DNA (positive or negative supercoiling, respectively) relatively to its relaxed state. In a DNA loop the number of times the double strands wrap around each other is fixed. So, in order to alleviate the torsional strain, caused by over- or under-winding, the DNA writhes up on itself. Supercoiling can be modelled by considering twisting rigidity potentials that can be defined by assigning an orientation to each one of the DNA polymer beads. One way of doing this is to decorate the surface of each bead with three patchy particles, as in the model of Ref.~[\onlinecite{Brackley2014}]. The positions of the patches are such that they establish three orthogonal unit vectors, which define the bead's reference frame. With this it is possible to define three angles that give the orientation of a bead with respect to its neighbour: one angle $\zeta$ that gives the orientation of the bead with respect to the polymer backbone, and two dihedral angles $\phi_1$ and $\phi_2$ that set the degree of twisting in the plane perpendicular to the backbone. Therefore, the orientation of the bead is kept aligned with the polymer backbone by defining the potential
\begin{eqnarray}\label{Superc-backbone}
U_{{\rm BB}}(\zeta)=K_{{\rm BB}}(1+\cos (\zeta)),
\end{eqnarray}
and the twisting rigidity of the polymer is modelled by assigning to both dihedral angles $\phi_{1,2}$ the energy
\begin{eqnarray}\label{Superc-twist}
U_{{\rm TW}}(\phi)=K_{{\rm TW}}(1+\cos (\phi+\phi_0)),
\end{eqnarray}
where $\phi_0$ is a phase related to the twist. More precisely, $\phi_0$ determines the number of beads involved in half a turn of the ribbon -- $\phi_0 = \pi + \frac{2\pi}{p}$, where $p$ is the pitch, i.e., the length scale corresponding to one full twist in equilibrium.

Proteins are also modelled, for simplicity, as spheres of diameter equal to $\sigma$, so that all beads in the simulation have the same size; in the Discussion and Conclusions section we also present selected results where we have considered a protein diameter of $2\sigma$ which matches more closely the size of a typical bacterial protein~\cite{phillips}. In the simulations, proteins interact sterically with each other via the WCA potential in Eq.~(\ref{V_WCA}). In this study we consider two kinds of proteins: non-DNA-binding and DNA-binding. The interaction between non-DNA-binding proteins and the DNA is purely repulsive and, again, described by the potential in Eq.~(\ref{V_WCA}). For the DNA-binding proteins, we assume that they are non-specifically DNA-binding, so can bind to any DNA bead. The attractive interaction is set by a Lennard-Jones (LJ) potential
\begin{eqnarray}\label{V_LJ_cut}
U_{\rm LJ shift}(r_{ij})= 
 \left\{
  \begin{array}{lr}
    U_{\rm LJ}(r_{ij}) - U_{\rm LJ}(r_{thr}) &  r_{ij}<r_{{\rm thr}} \\ 
    \qquad \qquad 0  & {\rm otherwise,}
  \end{array}
 \right.
\end{eqnarray}
\begin{eqnarray}\label{V_LJ}
{\rm where} \qquad U_{\rm LJ}(r)= 4\epsilon \left[ \left( \frac{d_{ij}}{r} \right)^{12} - \left( \frac{d_{ij}}{r} \right)^{6} \right]{\rm .}
\end{eqnarray}
The parameter $\epsilon$ controls the magnitude of the protein-DNA interaction, $d_{ij}=\sigma$ and $r_{{\rm thr}}$ is the range of the interaction. For the simulations where both kinds of proteins are present, $r_{{\rm thr}}=3.0\sigma$ and $\epsilon=2.5k_BT$, leading to a weak-moderate attractive protein-DNA interaction. For the simulations where only DNA-binding proteins are present, $r_{{\rm thr}}=1.5\sigma$ and we choose values for $\epsilon$ in the range $[2.0,5.0]$.

Note that we do not explicitly consider electrostatic interactions between proteins and/or DNA beads, again for simplicity, although these are charged in reality. This approximation is motivated by the fact that for a physiological 150 mM concentration of monovalent salt the Debye length is around $1$ nm \cite{Ando2010}, which is below the size of the proteins, so that electrostatic interactions are heavily screened in practice. We performed some simulations where we did consider charged particles (see Appendix B) and the results confirm that one can rely on this simplification.

The piston is modelled as a rigid sphere whose diameter is slightly larger than the confining cylindrical pore's diameter, in order to prevent particles from escaping around the sides. The interaction between the piston and the confined particles is purely repulsive and again described by the WCA potential [Eq.~(\ref{V_WCA})]. The piston moves under a uniform externally applied force, compressing the DNA and the proteins against the cylinder walls (see Fig.~\ref{pic1}).

We simulate the system by using Brownian dynamics (BD) via the Large-scale Atomic/Molecular Massively Parallel Simulator (LAMMPS) code in the BD mode. In other words, we use a molecular dynamics (MD) algorithm with a stochastic thermostat, which models the thermal fluctuations and viscosity of an implicit solvent. The system's temperature is kept constant at a value $T=1.0\equiv 300$ K. Each particle (DNA beads, proteins and piston) obeys a Langevin equation
\begin{eqnarray}
m \frac{d^2 \textbf{r}_{i}}{dt^2} = -\nabla_iV - \gamma \frac{d \textbf{r}_{i}}{dt}+ \sqrt{2k_BT \gamma}\boldsymbol\eta_{i} (t),
\label{motioneq}
\end{eqnarray}
where $\textbf{r}_{i}$ is the position of the centre of mass of the particle with mass $m=1$, $\gamma_i$ is the friction due to the solvent (typically $\gamma=0.5$) and $\boldsymbol\eta_i (t)$ is a vector representing random uncorrelated noise, such that
\begin{eqnarray}
&&\langle \eta_{i \alpha}(t) \rangle = 0, \nonumber \\ &&\langle \eta_{i \alpha}(t)\eta_{j \beta}(t') \rangle = \delta_{\alpha \beta}\delta(t-t')\delta_{ij}, 
\end{eqnarray}
where $\alpha$ and $\beta$ indicate Cartesian components, $\delta_{ij}$ and $\delta_{\alpha\beta}$ denote Kronecker's delta, and $\delta(t-t')$ denotes Dirac's delta function. Eq.~(\ref{motioneq}) is integrated with a constant time step $\Delta t = 0.01 \tau$, where $\tau$ is the simulation time unit, for a total of $3 \times 10^6$ time steps or more.

\subsection{\label{sec:units}Mapping between simulation and physical units}

In our simulations we use energy units of $k_BT$ (where $T=300$ K), length units of $\sigma=2.5$ nm (the size of a DNA bead that corresponds to the hydrated thickness of B-DNA), and mass units of $m_{\rm DNA}$, where the mass of a DNA bead $m_{\rm DNA}=1$. With this choice there is a natural simulation time unit $\tau=\sigma\sqrt{m/k_BT}$. Therefore, in our coarse-grained model, each DNA monomer represents $\sim 7.4$ bp of B-DNA (for which the distance between consecutive base pairs is $0.34$ nm).

In our physical system, there are two further timescales, beyond $\tau$. First, there is an inertial timescale, $\tau_{\rm in}=m/\gamma$, where $m$ is the mass of a DNA or protein bead, and  $\gamma$ is its friction. This quantity gives the time after which the velocity of the beads becomes uncorrelated. The second is the Brownian time, $\tau_B=\sigma^2/D_{\rm diff}$, which gives the (order of magnitude of the) time taken for a bead to diffuse across its own diameter, $\sigma$. The diffusion coefficient $D_{\rm diff}$ is set by the friction $\gamma$ through Einstein's formula $D_{\rm diff}=k_BT/\gamma$. We use $\gamma=2$, which leads to a Brownian timescale $\tau_B=2\tau=4\tau_{\rm in}$ (simulation units). Since we are interested in long time behaviours, i.e., of the order of several ms, and not in resolving fine details of the inertial collisions (which play no role in our overdamped system), in order to map time from simulation to physical units, we therefore match the Brownian timescale. As a sphere of diameter $2.5$ nm in an aqueous fluid (viscosity $1$ cP) has $\tau_B \approx 35.6$ ns, it follows that one simulation time unit $\tau$ corresponds to $17.8$ ns~\cite{Brackley2015}. Note that our choice of timescales means that we do not correctly describe the inertial dynamics before $\tau_B$, but this is not an issue for our purposes, as we are interested in timescales much exceeding this. At the same time, our choice of $\tau_B=2\tau=4\tau_i$ ensures that the dynamics in our simulations is overdamped, which is the physically relevant case for our system.

By using the mapping given above for energy, length and time we can map all of the other physical quantities we measure. For instance for the force, one simulation force unit is $k_BT/\sigma$, which corresponds to $1.64$ pN. 

As detailed above, the system studied here is composed of three types of beads: DNA monomers, non-DNA-binding proteins (crowders) and DNA-binding proteins. The number of beads is chosen so that the ratio of the number of proteins to the number of DNA beads is that found {\it in vivo}: The chromosome of {\it E. coli} consists of $\sim 4.6\times 10^6$ bp, which in our model corresponds to a polymer with $\sim 5\times10^5$ beads. In bacterial cells, the estimated total number of proteins is $M\sim10^6$, from which $\sim3\%$ are DNA-binding. This corresponds to a protein-DNA bead ratio of $2:1$. Since the computational cost associated with a $5\times10^5$ bead polymer is extremely high, we instead consider a smaller system with $N=1000$ beads. We then explore the influence of proteins in the system, by varying the protein number up to $M=2000$, which represents the true protein-DNA ratio. For simulations where both crowders and DNA-binding proteins are present, the number of DNA-binding proteins is $0.03M$.

\section{\label{sec:theory}Summary of Scaling Theory for Cylindrically Confined Polymers}

\subsection{\label{sec:EntrSpringTh}A Review of the Entropic Spring Theory}

According to the well known blob scaling concept in polymer physics, a self-avoiding polymer confined in a cylindrical pore can be seen as a linear chain of blobs~\cite{Brochard1977}. In~[\onlinecite{Jun2008}], this blob-scaling approach is used to derive an expression for the force required to compress a polymer confined in a cylindrical nanopore, assuming it behaves like an entropic spring. The ``renormalized'' free energy $\mathcal{F}$ for a confined polymer of $N$ beads is given by
\begin{eqnarray}
\beta \mathcal{F}(R,D) = A\frac{R^2}{(N/g)D^2}+B\frac{D(N/g)^2}{R},
\end{eqnarray}
where $R$ is the extension of the polymer along the axis of the cylindrical pore (see Fig.~\ref{pic1}), $\beta=1/k_BT$, A and B are constants, and g is the number of monomers inside a blob of diameter D. From this, a universal scaling relation for the (external) force-extension relation can be derived,
\begin{eqnarray}
D \beta f = D \frac{\partial}{\partial R} (\beta \mathcal{F}) = 2A \left(\frac{R}{R_0}\right) - B \left(\frac{R}{R_0}\right)^{-2},
\label{jun}
\end{eqnarray}
where $D$ can be seen as the confining cylinder's width. Here $R_0$ is the extension of the relaxed (zero force) confined polymer.
In the regime of weak deformations~\cite{Pelletier}, this reduces to
\begin{eqnarray}
f=\frac{k \ (k_BT)}{D} \left[ \left( \frac{R}{R_0} \right) - \left( \frac{R}{R_0} \right)^{-2} \right],
\label{pelletier}
\end{eqnarray}
where $k$ is the dimensionless polymer spring constant. 

In the regime of strong compression ($R/R_0 \ll 1$), the free energy in the blob-scaling formalism takes another form and the force-extension relation becomes~\cite{Jun2008}
\begin{eqnarray}
f \sim - \frac{k_BT}{D} \left( \frac{R}{R_0} \right)^{-9/4}.
\label{pelletier-strong}
\end{eqnarray} 

\subsection{\label{sec:PolExpDynamics}A Scaling Theory for the Polymer Expansion dynamics}

Concerning polymer dynamics, there are theoretical models that give an estimate of the evolution of the polymer extension in time, during polymer expansion under confinement. Taking $\mathcal{F}$ to be the confinement-deformation free energy, the polymer deformation force $f$ is given by $-\partial\mathcal{F}/\partial R$. The equation of motion for the polymer extension $R$ becomes~\cite{Jung2013}
\begin{eqnarray}
f+\frac{1}{2}\gamma_{polymer}\frac{dR}{dt}=0,
\label{R-motioneq}
\end{eqnarray}
where $\gamma_{polymer}$ is the polymer friction coefficient.
This equation can be solved given the initial condition $R(t=0)$ -- the extension of the compressed polymer just before it starts expanding. An explicit form for $f$ depends on the physical problem at hand. Following the approach in~[\onlinecite{Jun2008}], we use the force-extension relations in Eqs. (\ref{jun}-\ref{pelletier-strong}).
Taking Eq. (\ref{pelletier}) for the weak compression regime and substituting it into Eq. (\ref{R-motioneq}), one gets
\begin{eqnarray}
\frac{dR}{dt} = -\frac{2}{\gamma}\frac{k \ (k_BT)}{D} \left[ \left( \frac{R}{R_0} \right) - \left( \frac{R}{R_0} \right)^{-2} \right].
\end{eqnarray}
By defining $\tilde{R}=R/R_0$ and $A=\left(2 k (k_BT)\right)/\left(R_0 \gamma D\right)$, this becomes
\begin{eqnarray}
\frac{d\tilde{R}}{dt} = -A \left(\tilde{R} - \tilde{R}^{-2} \right) \ \ \Leftrightarrow \ \ \frac{\tilde{R}^2}{\tilde{R}^3-1}\frac{d\tilde{R}}{dt} = -A.
\end{eqnarray}
Taking into account that during the polymer expansion $\tilde{R}<1$, the integral of this equation leads to the solution
\begin{eqnarray}
&&\ln(1-\tilde{R}^3)=-3At+c \nonumber \\ \Leftrightarrow \ \ &&\tilde{R}(t)=\frac{R(t)}{R_0}= \left( 1-\tilde{c}e^{-3At} \right)^{1/3},
\end{eqnarray}
where $\tilde{c}=e^c$ and $c$ are constants that are defined by the initial condition $\tilde{R}(t=0)$: $\tilde{c}=1-\tilde{R}^3(t=0)$.

According to the solution $R(t)=R_0\left( 1-\tilde{c}e^{-3At} \right)^{1/3}$, there is an initial offset $R(t=0)$ corresponding to the extension of the polymer after being compressed by the piston. For intermediate times ($1\ \rm{ms} \lesssim t \lesssim 10\ \rm{ms}$), one can expand the solution about zero to get $R(t) \approx R_0\left( 1-\tilde{c} + 3A\tilde{c}t \right)^{1/3}$, which reveals the scaling $R(t)\sim t^{1/3}$. For late times the polymer relaxes completely reaching a saturation regime where the extension corresponds to the value at equilibrium in the absence of a compression force: $R(t \rightarrow \infty )=R_0$.

In the strong compression regime [Eq. (\ref{pelletier-strong})], the extension-time scaling relation for intermediate times becomes $R \sim t^{4/13}$.

These scaling relations are derived for a polymer confined in a cylindrical pore in the absence of other particles. However, the bacterial cellular environment is rich in proteins and so one can ask how the scaling would change in the presence of proteins either non-DNA-binding (crowders) or DNA-binding. Crowders have been shown to increase the viscosity of the medium~\cite{Toan2006}, which slows the polymer diffusion and hence the polymer expansion dynamics. Therefore one might expect a scaling relation for intermediate times $R \sim t^{\alpha}$ where the exponent $\alpha$ is smaller than that derived above in the absence of crowders. DNA-binding proteins with more than one binding site (such as the ones considered in the present study) bridge together DNA segments, reducing the polymer gyration radius and the extension of the confined polymer. This bridging effect increases the local polymer stiffness, reducing the effective polymer elasticity. Therefore, one may also expect a slow-down effect of the polymer expansion in the presence of such proteins and a further decrease of the exponent $\alpha$.

\section{Results}

\subsection{\label{sec:setup}Set-up of the simulation}

\begin{figure}
\begin{center}
\includegraphics[width=9.cm]{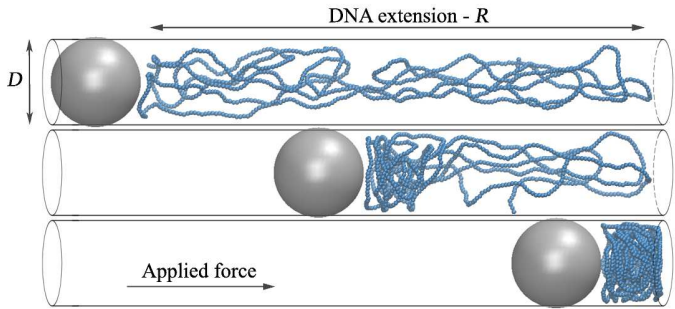}
\end{center}
\caption{\label{pic1}Sketch of our model for a DNA molecule, confined in a cylindrical channel, being compressed by a piston subjected to an external force. The DNA extension is measured as the largest distance between two DNA beads, along the confining cylinder's axis. To build our force-extension curves, we recorded the DNA extension after equilibration, for each value of the force (an example is given in the snapshot in the bottom picture). The two top pictures correspond to transient configurations during compression.}
\end{figure}

\begin{figure}
\begin{center}
\includegraphics[width=8.8cm]{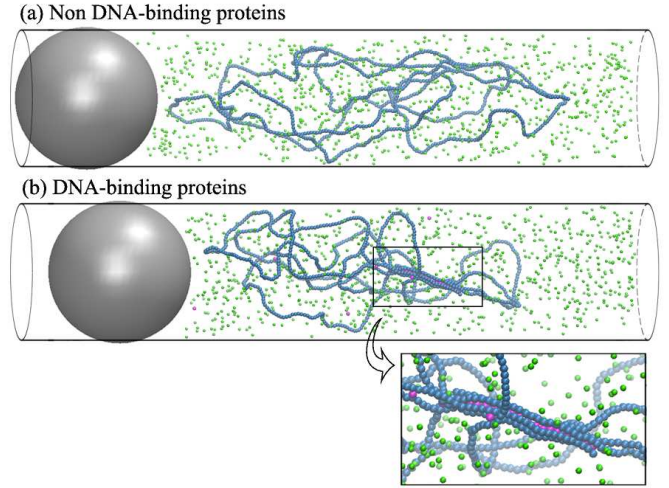}
\end{center}
\caption{\label{pic2}Sketch of our model for a DNA molecule in the presence of (a) non-DNA-binding proteins (green spheres) and (b) DNA-binding proteins (magenta spheres). The DNA-binding proteins cluster, inducing a local tubular folding of DNA (see inset).}
\end{figure}

First, we discuss our simulation geometry and set-up (see Fig.~\ref{pic1}). We considered two cases: DNA-only simulations (Fig.~\ref{pic1}), and a DNA molecule interacting with proteins (both DNA-binding and non-DNA-binding, Fig.~\ref{pic2}(a) and Fig.~\ref{pic2}(b) respectively). Note that the two top pictures in Figure~\ref{pic1} correspond to transient configurations during compression of the DNA molecule, depicting the time evolution of the simulation. The bottom picture corresponds to the fully equilibrated system.

To simulate DNA, we considered a model linear DNA molecule of $N=1000$ beads, within cylinders of two different diameters ($D=40 \sigma$ and $D=20 \sigma$, the latter case is shown in Fig.~\ref{pic1}(a)). Results for the two diameters are very similar, hence we show below only those for $D= 40 \sigma$. For each configuration we measure the DNA extension along the longitudinal direction of the cylinder, defined as the maximum distance between any two beads in the chain -- we denote this quantity as $R$. We then take the average of $R$ over 10 configurations. The radius of gyration of the unconfined DNA (in the absence of the cylinder) was estimated to be $R_g\sim 62.2\sigma$, while its extension within the cylinder in the absence of the piston was measured as $R_0\sim 152.2\sigma$. Therefore, we work in the semi-dilute regime which is realistic for the bacterial chromosome. Although in {\it E. coli} the DNA is circular, the scaling theory and previous work suggests that the main contribution are given by polymer confinement (entropy and free energy loss). Hence we expect the polymer topology to be not too important -- the simulations with circular DNA described in Section~\ref{sec:sim-supercoiling} further support this view.

For the DNA-protein simulations, we considered two situations. In the first (Fig.~\ref{pic2}(a)), we only included non-DNA-binding proteins (green spheres in Fig.~\ref{pic2}(a)), which are the majority {\it in vivo}: for instance, in {\it E. coli} there are about $10^6$ proteins, of which only $3\%$ are estimated to be DNA-binding. As the bacterial cell volume of {\it E. coli} is about $6\mu$m$^3$, and the average diameter of a bacterial protein is $5$ nm~\cite{phillips}, the volume fraction occupied by all bacterial proteins is a few \%, which is larger than the volume occupied by DNA. In our simulations, we typically considered $M=2000$ proteins, modelled as spheres (matching the ratio found {\it in vivo} between number of proteins and number of ``DNA beads''). In the second situation, we included a fraction of DNA-binding proteins (magenta spheres in Fig.~\ref{pic2}(b)); we simulated non-specific binding, as is the case, to a first approximation, for the histone-like bacterial H-NS protein~\cite{HNS}, or for the DPS protein, which is activated following bacterial starvation~\cite{DPS}. While we include no direct protein-protein attractive interaction, DNA-binding proteins naturally cluster, through the so-called ``bridging-induced attraction'' described previously in Refs.~[\onlinecite{Brackley2013,Brackley2016}]. This attraction is associated with a simple positive feedback loop: proteins bind to the DNA in multiple places forming bridges, this increases the local concentration of DNA, which in turn recruits further proteins, etc., ultimately resulting, for the protein size considered here, in the formation of elongated protein clusters associated with a local tubular folding of DNA where the protein-associated segments are parallel to each other. Ref.~[\onlinecite{Brackley2013}] shows that larger proteins lead to DNA wrapping around them. However this is not realistic for H-NS which is thought to form linear clusters~\cite{Brackley2013,HNS}, as in Figure~\ref{pic2}.

Figures~\ref{pic1} and \ref{pic2} show examples of compression simulations, where different forces were applied to the piston, and the resulting DNA conformations were analysed. We have also performed further simulations, as follows. First, we considered compression simulations with modified protein density. Second, we simulated the dynamics of entropic expansion, by first compressing the DNA with a large force, allowing the system to reach equilibrium, and subsequently removing the piston. Third, we have varied the strength of the interaction between DNA-binding proteins and DNA. Finally, we have also considered the compression of supercoiled DNA loops (as opposed to the linear DNA shown in Figs.~\ref{pic1} and \ref{pic2}).

\subsection{\label{sec:sim-compressDNA}Excluded volume effects lead to quantitative deviations from the predicted entropic spring response in DNA-only simulations}

\begin{figure*}
\begin{center}
\includegraphics[width=16.cm]{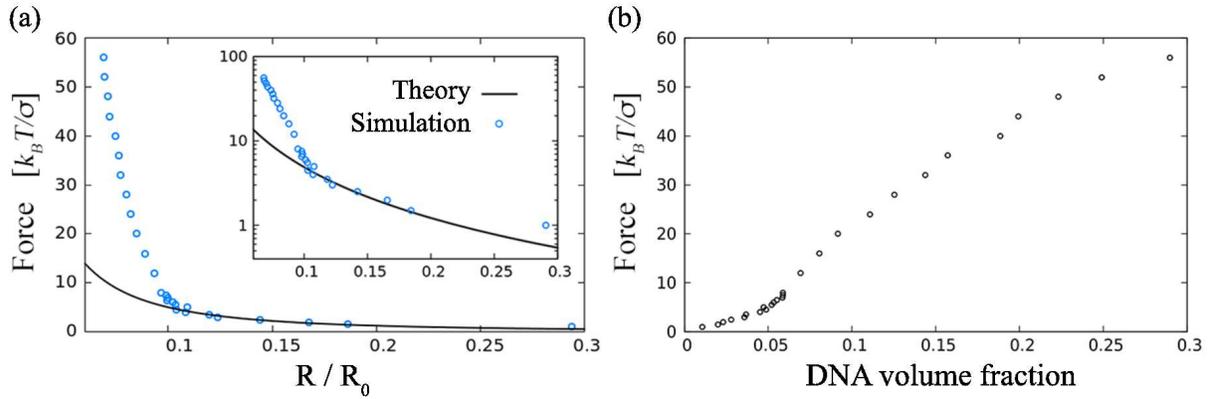}
\end{center}
\caption{\label{pic3}(a) Comparison of the results for the compression force as a function of the normalized DNA extension in the absence of proteins, from the entropic spring theory of Ref.~[\onlinecite{Pelletier}] (full line) and our numerical model (blue circles). Parameters were as specified in the Section~\ref{sec:MaterialsandMethods}, apart from $D=40\sigma$. 
For $R/R_0>0.1$, the theory agrees with the model in that there is a strong dependence of the DNA extension on the compression force. For $R/R_0<0.1$, the theory deviates from our numerical estimate of the entropic force, suggesting that, in the high compression regime, the assumption done in the theory of an infinitesimally thin polymer, where excluded volume effects can be disregarded, breaks down. The inset shows the results for the ``Force'' axis in logarithmic scale. (b) Compression force as a function of the DNA volume fraction. As the DNA volume fraction increases, an increasing force is needed to further compress the DNA. The fact that the volume fraction reaches almost $0.3$ for the tightest compression suggests that close packing and many body interactions are the most relevant contributions to the divergence of our numerical estimates from the theory in the high compression regime observed in plot (a).}
\end{figure*}

We begin by studying the entropic compression in DNA-only simulations (Fig.~\ref{pic3}(a)). We find two regimes. First, for $R/R_0>0.1$ the entropic spring theory revised in Section~\ref{sec:EntrSpringTh} (which is also the theory used in Ref.~[\onlinecite{Pelletier}]) works well, and we observe a very sensitive dependence of the extension on compression force: the extension reduces to 10\% of its free value, $R_0$, for a force of just $\sim 3$ simulation units (corresponding to about 5 pN -- see mapping in Sec.~\ref{sec:units}). Second, for $R/R_0<0.1$, we find a sharp deviation from the theory: our numerical estimate of the entropic force is more than an order of magnitude larger than predicted. The origin of this deviation is in the assumption in the scaling theory of an infinitesimally thin polymer: clearly, this is not applicable to the tightly confined regime which we reach by the end of our simulation, where the DNA segments are forced into close contact, so that steric interactions dominate. Figure~\ref{pic3}(b) further suggests that close packing and many body interactions are the main contributions to this strong steric repulsion in the tightly confined regime, as it can be seen from the fact that the volume fraction reaches a large value ($\sim 0.3$) for the largest compression force.

This result suggests that steric repulsion is an important factor to consider when estimating the entropic force resisting compression. At the same time, we note that the effect found here (a 10-fold increase) is an overestimate of the correction needed for a real bacterial chromosome. This is due to the scaled down dimensions of the DNA we use: the same value of $R/R_0$ corresponds to a much denser volume fraction of DNA in our simulations with respect to the experiments in Ref.~[\onlinecite{Pelletier}]. 

\subsection{\label{sec:sim-compressProt}Non-DNA-binding proteins greatly increase the entropic force and pressure exerted on the piston}

Next, we analyse the effect of bacterial proteins on the entropic elasticity of the system. The osmotic pressure of proteins in the cytosol can be estimated as $M k_BT/V$, where $M$ is the number of proteins, and $V$ is the confinement volume; for the case of {\it E. coli} this is $\sim$0.01 atm, which is larger than, or at least of the same order of, the pressures recorded in the experiment in Ref.~[\onlinecite{Pelletier}].

\begin{figure}
\begin{center}
\includegraphics[width=8.4cm]{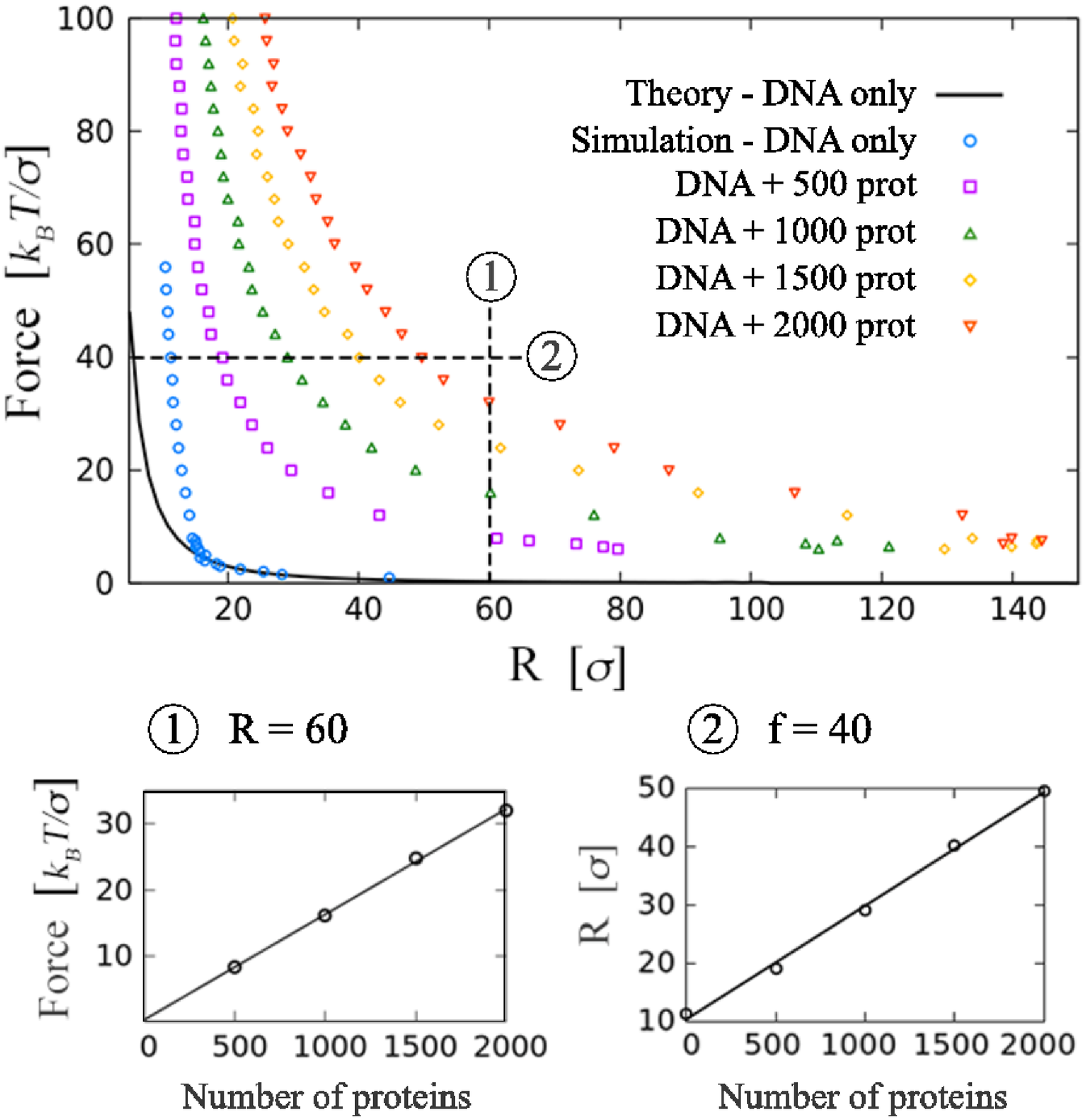}
\end{center}
\caption{\label{pic4}Comparison of the results for the compression force as a function of the DNA extension, from the DNA-only simulations (blue circles) and the simulations in the presence of non-DNA-binding proteins. The protein crowding effects were studied by examining systems with different numbers of proteins: $M=500$ (magenta squares), $M=1000$ (green triangles), $M=1500$ (yellow diamonds) and $M=2000$ (orange inverted triangles). The protein osmotic contribution leads to a large increase in the compression force, sometimes of several orders of magnitude. For a given value of $R$, the protein contribution is linear in the number of proteins (inset 1) -- fit: $f=0.0160M+0.2933$. For a given compression force, the DNA extension also increases linearly with the increasing number of proteins (inset 2) -- fit: $f=0.0195M+10.4113$.}
\end{figure}

We show in Figure~\ref{pic4} how the presence of (non-DNA-binding) proteins (modelled as spheres which interact both with each other and with the DNA solely via excluded volume) affects the force-extension curves. We find that proteins make an important contribution, especially for moderate compression (relatively large values of $R$): in this regime, the force required for a given extension is orders of magnitude larger than in the case where the proteins are not included. 
More specifically, the force is approximately linear in the number of proteins $M$, as expected for an osmotic contribution (see Fig.~\ref{pic4}, inset 1). Interestingly, for intermediate forces, we find that the extension recorded at a given force also depends linearly on $M$ (Fig. ~\ref{pic4}, inset 2). 

\begin{figure}
\begin{center}
\includegraphics[width=8.4cm]{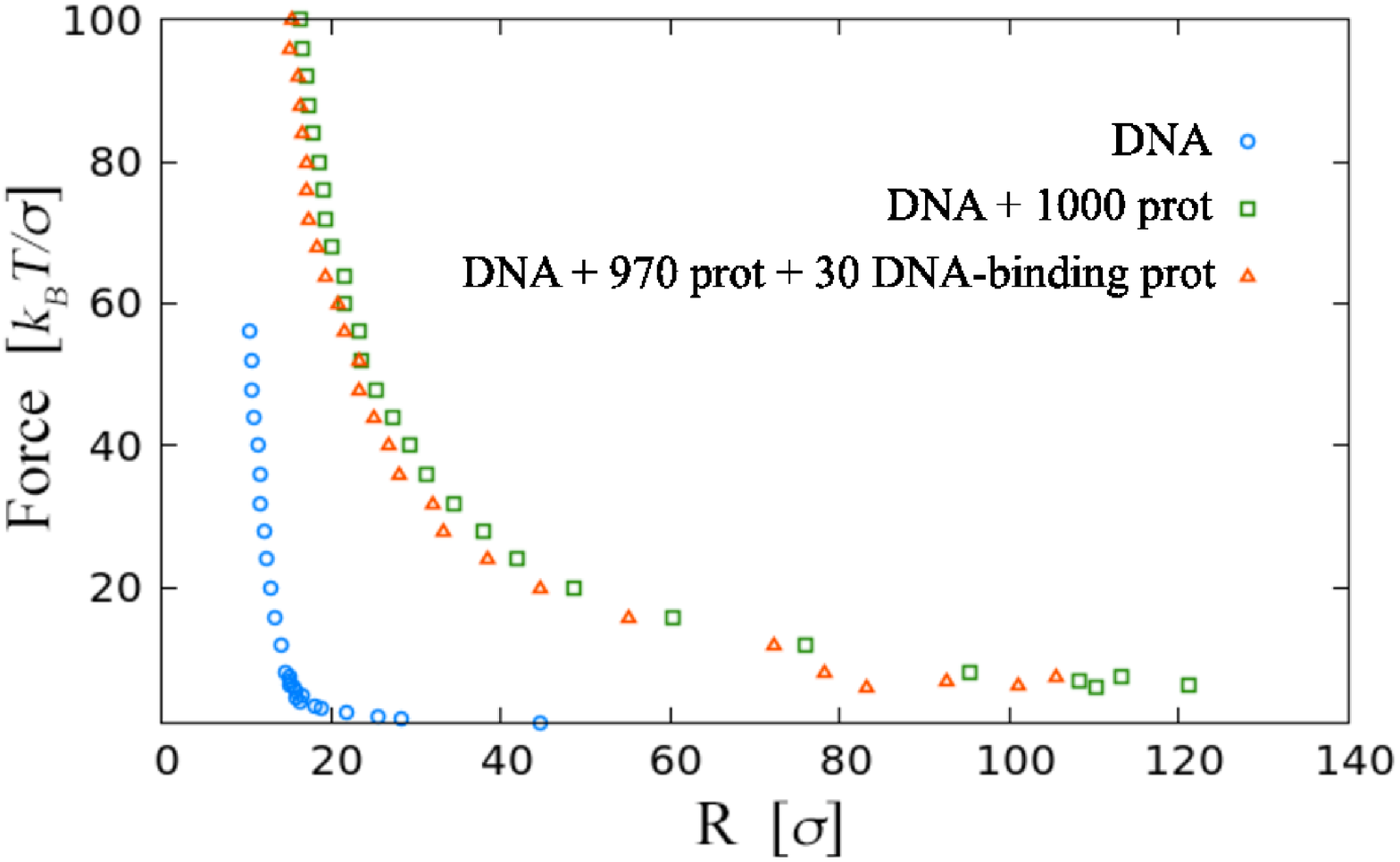}
\end{center}
\caption{\label{pic5}Compression force as a function of the DNA extension for DNA-only simulations (blue circles), simulations in the presence of non-DNA-binding proteins (green squares) and simulations in the presence of both types of proteins (orange triangles): $3\%$ of the proteins are DNA-binding. The DNA-binding proteins lead to formation of DNA clusters, hence compacting the DNA, which yields a decrease in the compression force. The effect in the osmotic pressure due to DNA-binding proteins is, however, small.}
\end{figure}

We next address the role of DNA-binding proteins. These tend to compact the DNA, and should lead to a change in the extensional elasticity. Mobile cross-links, or slip-links, have previously been shown to strongly modify the force-extension curves of polymers in stretching experiments~\cite{Kardar}; for this reason, DNA-binding proteins were proposed in Ref.~[\onlinecite{Pelletier}] as a possible explanation for the quantitative significant discrepancy between the experiments and the entropic spring theory in compression experiments. Therefore we ask to what extent the presence of DNA-associating proteins affect our results. Fig.~\ref{pic5} shows that these effects are minor in the simulations. This is consistent with our previous finding that osmotic forces from non-DNA-binding proteins are more important quantitatively than DNA entropic forces: DNA-binding proteins only affect the polymer response, and hence do not have much bearing on the overall curve. We should stress here that the DNA-binding proteins we consider form clusters, as could be the case for H-NS~\cite{Brackley2013,HNS}, or DPS~\cite{DPS}; while the dynamic crosslinks invoked in Ref.~[\onlinecite{Pelletier}] will in practice interact entropically (see~[\onlinecite{Kardar2}]), their collective behaviour may be different. 

In summary, we find that non-DNA-binding proteins exert a significant osmotic pressure on the piston, and this is much larger than the force exerted by the DNA. Furthermore, any reduction in the force exerted by the DNA due to DNA-binding proteins is dwarfed by the contribution of the non-DNA-binding proteins. Our estimates suggest that even in the experimental situation the presence of non-DNA-binding proteins could significantly affect the force measured via the set-up used in Ref.~[\onlinecite{Pelletier}]; therefore it would be of interest to compare in more detail those experiments with {\it in vitro} compression experiments with different size DNA and different protein environments.

\subsection{\label{sec:sim-expansion}The expansion dynamics of DNA depends on macromolecular crowding and DNA-protein interactions}

\begin{figure*}
\begin{center}
\includegraphics[width=16.cm]{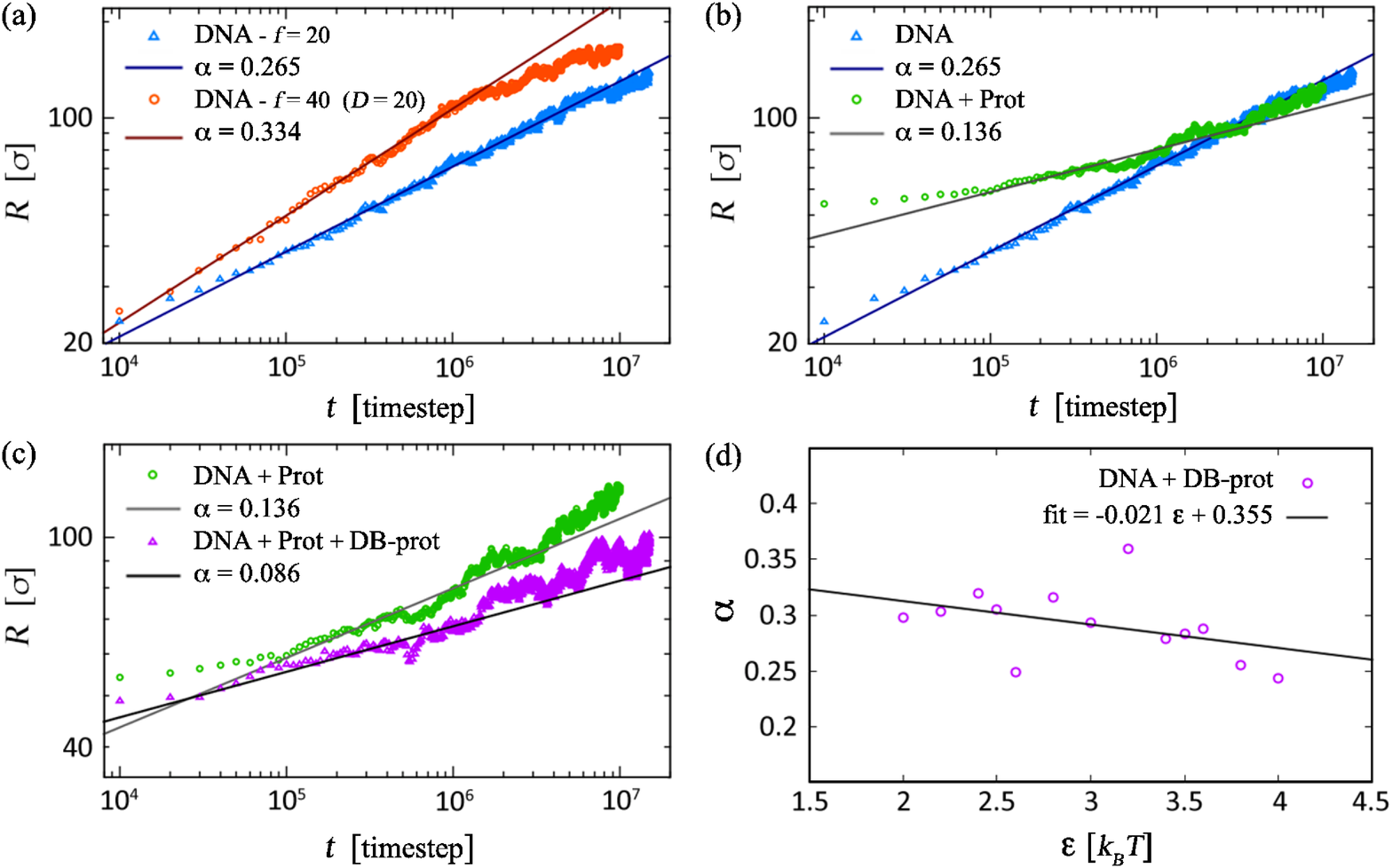}
\end{center}
\caption{\label{pic6}Measured dynamical exponents during DNA expansion (in all cases, curves correspond to averages over 10 runs). (a) Comparison of the exponents measured for DNA-only simulations for different initial compression forces: $f=20\ \epsilon/\sigma$ (blue triangles) and $f=40\ \epsilon/\sigma$ (orange circles). The bigger the compression force, and hence the more compressed the initial DNA configuration, the bigger the apparent value of the exponent. The lines were fit for the ranges of times $[10^4:8\times10^6]$ ($f=20\ \epsilon/\sigma$) and $[10^4:1.5\times10^6]$ ($f=40\ \epsilon/\sigma$). (b) Comparison of the exponents measured for DNA-only simulations (blue triangles) and simulations in the presence of $1000$ non-DNA-binding proteins (green circles). The protein crowding leads to a decrease in the exponent. The line for the case with DNA and non-DNA-binding proteins was fit for the range of times $[10^4:3.3\times10^6]$. (c) Comparison of the exponents measured for simulations in the presence of $1000$ non-DNA-binding proteins (green circles) and in the presence of $970$ non-DNA-binding and $30$ DNA-binding proteins (magenta triangles). The exponent decreases in the presence of DNA-binding proteins, suggesting that the formation of protein-induced DNA clusters slows down the DNA expansion. The line for the case with DNA, non-DNA-binding proteins and DNA-binding proteins was fit for the range of times $[10^4:1.6\times10^6]$. (d) Measured exponents for simulations with DNA and $100$ DNA-binding proteins, but no non-DNA-binding proteins, as a function of the DNA-protein binding strength. The exponents show a weak tendency to decrease with increasing interaction strength, supporting the observation that DNA clustering slows down DNA expansion. All simulations started from an initial DNA configuration obtained for a compression force of $20$, except for the case in the top-left plot. The ranges of times, for which the expansion curves were fit, were chosen to take into account just the first expansion regime.}
\end{figure*}

We next examine the dynamical behaviour of a DNA molecule (with or without proteins), where the DNA is first compressed under a strong force, and then let free to expand after the piston is removed. Figure~\ref{pic6} shows how the extension $R$ increases in time following the piston removal. We note that the curves for a single realisation are very noisy, so in Figure~\ref{pic6} each curve (or point) corresponds to an average over 10 independent runs. 

First, for the DNA-only case, one may expect a scaling behaviour for $R$ with $t^{4/13}$ (for self-avoiding chains, see Sec.~\ref{sec:PolExpDynamics} and Ref.~[\onlinecite{Jung2013}]). Our results typically show a similar, but on average slightly smaller, exponent: the values are also consistent with those found numerically in Ref.~[\onlinecite{Jung2013}]. Our simulations show a different apparent exponent for different compression force (Fig.~\ref{pic6}(a)): this may either point to some dependence on the initial condition, or to a wide variability of the exponent (which is apparent from Fig.~\ref{pic6}(d) and further discussed below).

Second, we consider the case with proteins. The expansion dynamics is much slower (i.e., the apparent exponent is smaller, Fig.~\ref{pic6}(b)) in the presence of non-DNA-binding proteins: this is because the proteins create a crowded environment which hampers DNA unfolding. At the same time, the DNA extension is larger at $t=0$, for the same initial force, because of the osmotic pressure of the proteins which opposes the compression force from the piston; as a result the value of $R(t)$ is always larger when proteins are present in our simulations. For late times, the proteins diffuse away from the DNA and become dilute, so the effect of crowding diminishes. Hence for large $t$ the apparent exponents for our curves with and without proteins become approximately equal.

DNA-binding proteins lead to even slower progress in the DNA expansion (Fig.~\ref{pic6}(c)), especially at early times. Therefore, DNA-binding proteins have a much more detectable signature in the expansion dynamics, with respect to the minor contribution noted in the force-extension curves in Figure~\ref{pic5} where DNA-binding proteins were absent. To characterise this effect more in detail, we performed simulations with {\it only} DNA-binding proteins (Fig.~\ref{pic6}(d)): the exponents found in this case should then be compared with those for DNA-only simulations (Fig.~\ref{pic6}(a)). These results confirm that when proteins bind to the DNA they may reduce the apparent exponent; however the effect is subtle, and there is a large stochastic element in the dynamical curves (as each of the values in Fig.~\ref{pic6}(d) is computed by averaging 10 different expansion runs). 

\subsection{\label{sec:sim-popoff}``Popping-off'' dynamics with DNA-binding proteins}

In Figure~\ref{pic6} (bottom two panels), we analysed the effect of DNA-binding proteins on the dynamics of DNA expansion. By examining the trajectories of the system as a function of DNA-protein interaction strength, we observed some further interesting phenomena in the kinetics of the system, which we now describe. 

If the DNA-protein interaction is weak ($\epsilon=2 k_BT$), proteins only transiently bind to DNA. Since increased concentration favours the bound state (the entropy loss upon binding is smaller), more proteins bind upon compression; however they detach immediately after the piston is released (see Fig.~\ref{pic7}, left column, and Fig.~\ref{pic7}, blue line in bottom panel). As a result, the DNA responds elastically as a protein-free polymer -- this is consistent with our finding that the apparent exponent measured for a low DNA-protein interaction strength $\epsilon$, in the simulations with only DNA-binding proteins, is comparable with the exponent measured for the DNA-only simulations (see Fig.~\ref{pic6}). 

Some interesting dynamics occur if we choose a larger DNA-protein interaction strength, so as to promote more long-lived binding (see Fig.~\ref{pic7}, middle column). In this situation, the interaction strength is such that it favours long-lived binding under compression -- in this case, essentially all proteins are bound at all times, and they locally compact the DNA into a toroidal structure, resembling that of DNA within bacteriophages~\footnote{In a real bacterial DNA, proteins like H-NS would not bind the whole genome so the 3D structure would likely be different upon compression; however, we expect similar dynamics to occur there as well.}. When the piston is removed, the translational entropy of the proteins in the unbound state increases dramatically -- as they could now occupy any region of the cylindrical domain; however the bound state is still metastable, and it takes a relatively long time for the proteins to detach. Over time, proteins ``pop-off'', typically one-by-one, from the collapsed DNA, and the total energy of the polymer-and-protein system (which is approximately proportional to the number of bound proteins) decreases in magnitude linearly with time. Therefore, the ``popping-off'' time should increase linearly with number of DNA-binding proteins in the system, so could easily be observable in experiments with bacterial DNA (recall there are an estimated $3\times 10^4$ DNA-binding proteins in the bacterial nucleoid). Intriguingly, while the protein kinetics are completely different, the popping-off leaves little detectable signature in the apparent exponents recorded in Figure~\ref{pic6}. Finally, the popping-off requires tuning of the interaction, because if this becomes too large, proteins bind to the DNA permanently (at least within our simulation time), and the popping-off kinetics can no longer be observed -- the energy now does not appreciably depend on time (see Fig.~\ref{pic7}, bottom panel, green curve). 

\begin{figure}
\raggedleft
\includegraphics[width=9.2cm]{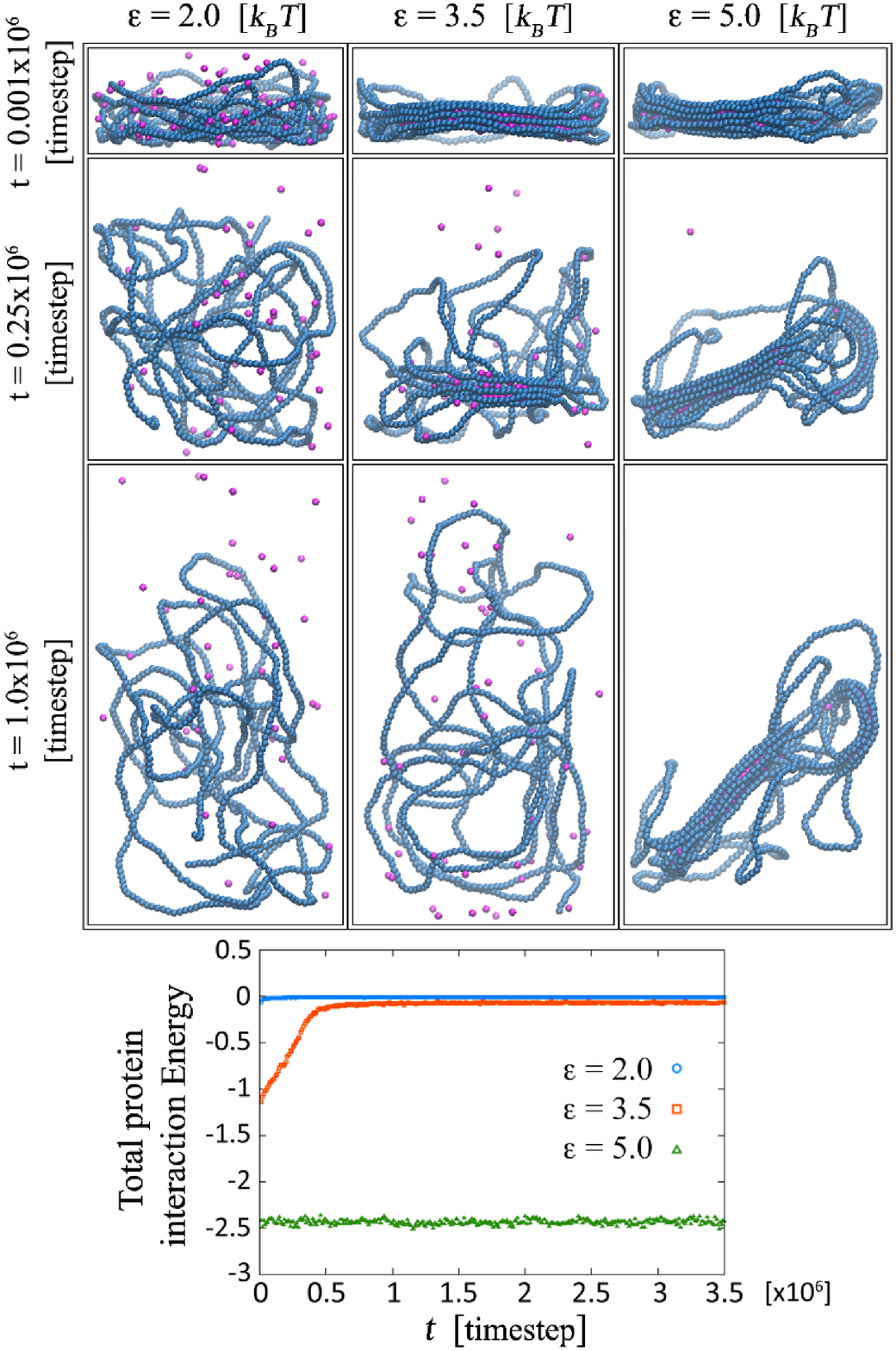}
\label{pic7}
\caption{\label{pic7}Protein popping-off during DNA free expansion, for simulations with DNA and $100$ DNA-binding proteins only. The bottom plot shows the overall pair energy of the system as a function of time. This quantity is negative and approximately proportional to the number of proteins bound to the DNA at a given time. For low DNA-protein interaction strength ($\epsilon=2.0$), the proteins do not stick to the DNA, as confirmed in the energy-time plot: the pair energy remains zero over time. For large DNA-protein interaction strength ($\epsilon=5.0$), the proteins remain permanently bound to the DNA during expansion, and the pair energy remains approximately constant over time. For an intermediate interaction strength ($\epsilon=3.5$), after removing the piston some proteins remain bound to the DNA, but eventually pop off. The pair energy first decreases linearly in time and then tends to zero asymptotically.}
\end{figure}

\subsection{\label{sec:sim-supercoiling}Compression curves and dynamics for supercoiled DNA}

Thus far, as previously highlighted, we have described the results obtained for linear DNA. Within bacteria, DNA is circular and negatively supercoiled (see Sec.~\ref{sec:introduction}); it is therefore of interest to ask what the effect of supercoiling is on our results.

In order to address this issue we consider the model introduced in Ref.~[\onlinecite{Brackley2014}] and described in Section~\ref{sec:model}. We consider here the same bending persistence length used previously in this work -- $\sim 50$ nm, appropriate for  B-DNA --  and a twisting rigidity of $\sim 75$ nm, which is in line with experimental values~\cite{twistDNA}. In DNA there is a complete twist every $10.4$ base pairs (corresponding to approximately $3.5$ nm). To set these parameters, we chose a bead size $\sigma = 3.4$ nm, $K_{\rm BEND} = 15.2\epsilon$, $K_{\rm TW} = 10\epsilon$ and $K_{{\rm BB}} = 90\epsilon$. We considered two situations. In the first case the dihedral phase $\phi_0$ is equal to $\pi$ giving a ribbon with no additional twist ($p = \infty$). In the second case $\phi_0 = \pi + \frac{2\pi}{10}$ giving a ribbon that is undertwisted by a full turn every $20$ beads i.e. every $208$ base pairs. This second case corresponds to supercoiled DNA, with a supercoiling density of $-0.05$ (note that in our model there is symmetry between negative and positive supercoiling because DNA denaturation is disallowed). In both cases, and as previously for the linear DNA, we considered a chain with $N=1000$ backbone beads (each with its associated patches). 

The chain was first relaxed within a cylinder of radius $R=20\sigma$ and length $400\sigma$, verifying that the linking number of the circular ribbon was constant and approximately equal to 0 and 50 (in absolute value) for the case of torsionally relaxed and supercoiled DNA respectively (data not shown).

\begin{figure*}
\begin{center}
\includegraphics[width=16.cm]{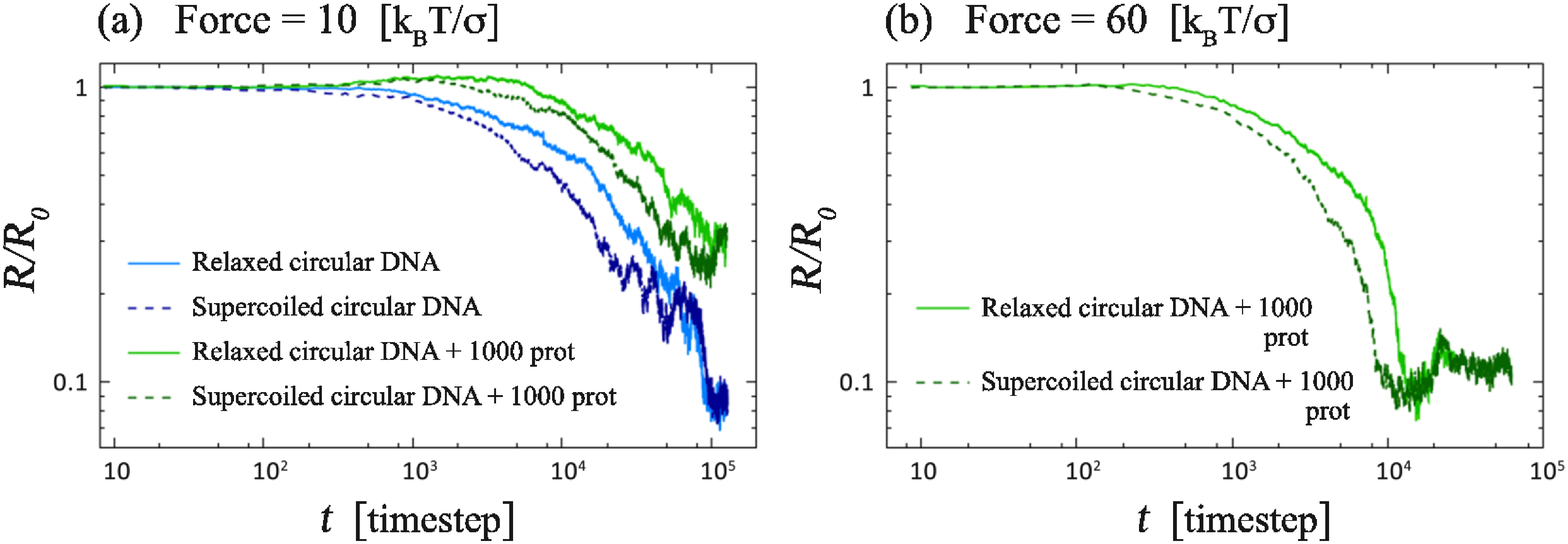}
\end{center}
\caption{\label{pic8}Compression dynamics under a force (a) $f=10$ and (b) $f=60$. The ratio $R/R_0$ is reported as a function of time for both the torsionally relaxed (full line) and the supercoiled (dashed line) DNA. Green and blue curves refer to the case in which $M=1000$ non-binding proteins are respectively present or absent in the system.
}
\end{figure*}

In Figure~\ref{pic8}(b) we report the time evolution of the longitudinal extension of the DNA, $R$, relative to its value $R_0$ in the relaxed state -- the compression was performed in the presence of $M=1000$ non-binding proteins, and by applying a constant force $f=60$ (in simulation units). For this value of $f$  the DNA-proteins system is already strongly compressed ($R/R_0 \sim 0.1$) independently on the degree of twisting.
While the final value of $R$ is very similar for the torsionally relaxed and the supercoiled DNA, the dynamics is, quite intriguingly, rather different, as the supercoiled DNA shows a softer response to compression. 
For mild compression $(f=10)$ (see Fig.~\ref{pic8}(a)) the situation is similar although, as expected, the final state is less compressed than the $f=60$ case ($R/R_0 > 0.3$). 
These conclusions also hold in the absence of non-binding proteins (blue curves in Fig.~\ref{pic8}(a)), however in that case the final value of $R$ is significantly lower ($R/R_0 \sim 0.07)$: as found previously with linear DNA, non-binding proteins contribute significantly to the force opposing compression by the piston. 

\begin{figure}
\begin{center}
\includegraphics[width=8.6cm]{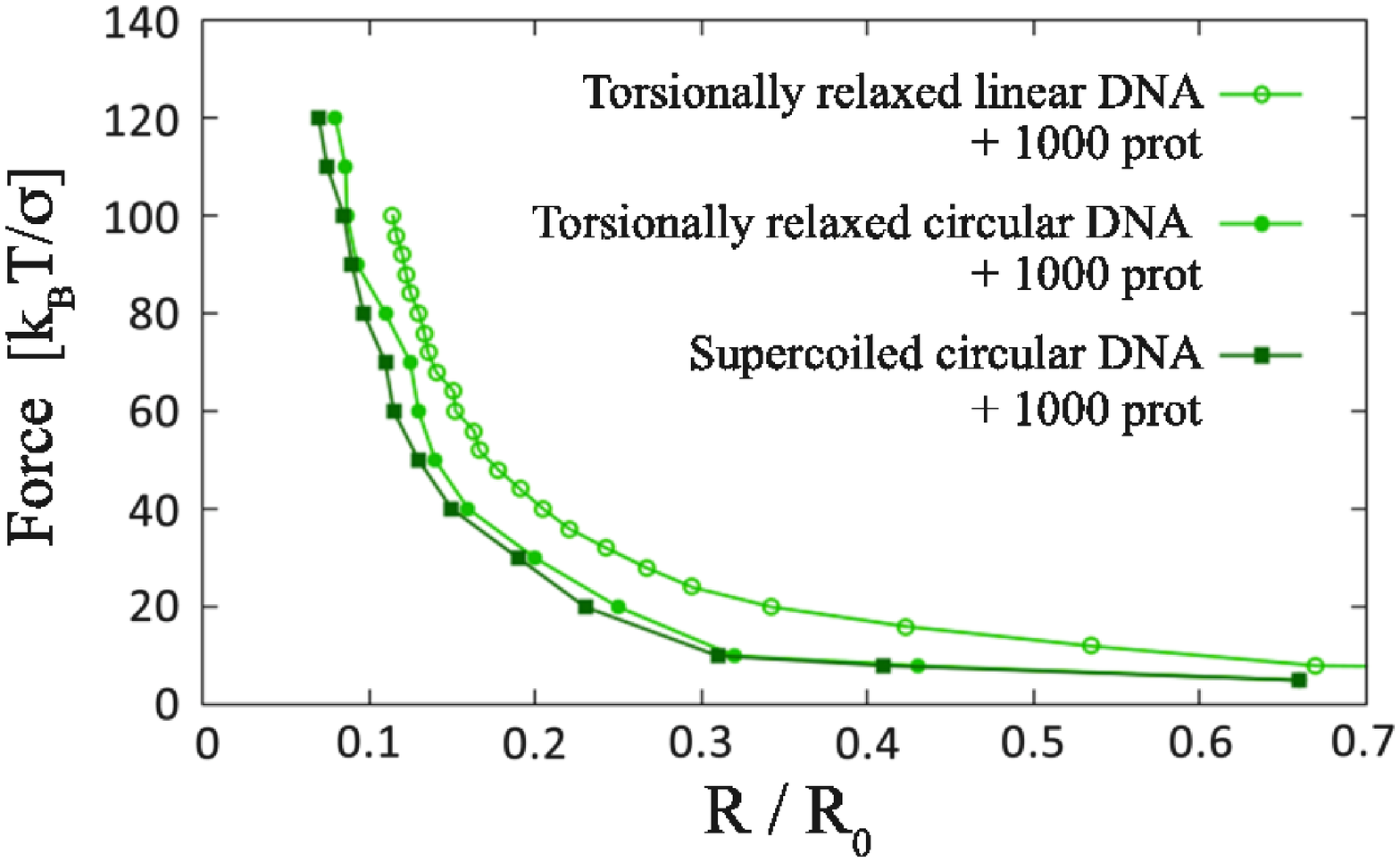}
\end{center}
\caption{\label{pic9}Compression force as a function of the extension ratio $R/R_0$ for torsionally relaxed (green circles) and supercoiled (dark-green squares) DNA. The curves with hollow and full points refer to the case of linear and circular DNA, respectively. The compression has been performed in presence of $M=1000$ non-binding proteins.}
\end{figure}

By considering different values of the force and looking at the equilibrium compressed state, we further computed the compression force as a function of the ratio $R/R_0$, for the torsionally relaxed and supercoiled DNA with $M=1000$ non-binding proteins. The results are reported in Figure~\ref{pic9}: both curves are quite similar, and, remarkably, relatively close to the linear case as well. 

Supercoiling or, in other words, the local writhing of the DNA polymer has an effect similar to that of DNA-binding proteins (described in Secs.~\ref{sec:sim-compressProt} and \ref{sec:sim-expansion}) -- it leads to a change in the extensional elasticity and hence a softer response to compression. Like in the case of DNA-binding proteins, the effect of supercoiling in the force-compression curves is very small, being more detectable in the dynamics curves.

\section{\label{sec:DiscussConclusion}Discussion and conclusions}

In summary, in this work we have presented Brownian dynamics simulations of the compression and expansion dynamics of a DNA molecule, modelled as a self-avoiding linear or looped polymer, interacting with an ensemble of non-DNA-binding and DNA-binding proteins. Previous work (see, among other, Refs.~[\onlinecite{deVries2012,Joyeux2015,Kim2015,Shendruk2014,Brackley2013}]) had clearly demonstrated the importance of proteins on the thermodynamic conformations of bacterial DNA: non-DNA-binding proteins create macromolecular crowding which can promote global collapse of the bacterial chromosomes, while DNA-binding proteins such as H-NS provide local compaction. The novelty of our current work is that, by addressing a set-up which is directly relevant to single molecule experiments probing the entropic elasticity of bacterial DNA, we establish that the the inclusion of proteins further leads to important effects on the force-extension curves recorded upon compression, and also on the expansion {\it dynamics} of the polymer.

First, we presented some DNA-only simulations, where there are no proteins as a reference case. We have shown that for weak to intermediate compression our results confirm the entropic spring theory of Ref.~[\onlinecite{Pelletier}], which predicts an abrupt decrease in elongation with compression. For large compression ($R/R_0<0.1$ in our simulations), however, excluded volume interaction create a strong deviation from the theory.

Second, we have found that the osmotic pressure of non-DNA-binding proteins can dwarf the entropic pressure of the spring-like polymer during compression. In our simulations, we have considered $N=1000$ polymer ``beads'', and up to $M=2000$ non-DNA-binding protein ``beads''; although these numbers are comparable, the effect of proteins can be orders of magnitude larger in the compression curves. With respect to these thermodynamic curves, DNA-binding proteins affect the DNA elasticity, hence only have a minor effect overall.

It is interesting to ask whether osmotic pressure from unbound proteins can account for the high force ($\sim 100$ pN) observed experimentally in the compression curves in Ref.~[\onlinecite{Pelletier}]. As mentioned by the authors, while the curve overall can be fitted with the entropic spring theory, the numerical value which is expected of a self-avoiding polymer would be significantly smaller than observed: we note that even a small fraction of proteins could lead to an osmotic force which might account for this. 

Third, we have explored the impact of proteins (again, non-DNA-binding and DNA-binding), on the expansion dynamics starting from a compressed DNA-and-protein system and following the removal of the force exerted by the piston. We found that the DNA-only simulations lead to a scaling behaviour for the extension $R$ as a function of time which is consistent with previous work in the literature~\cite{Jung2013}. The crowding introduced by non-DNA-binding proteins leads to a much slower dynamics, and to a much decreased apparent exponent. An interesting observation is that, when simulating DNA with DNA-binding-proteins only, tuning of the DNA-protein interaction leads to a popping-off kinetics during DNA expansion, where proteins are metastably bound to the DNA and detach one-by-one after the volume at their disposal increases.

Fourth, we have also performed more realistic simulations for bacterial DNA, where we considered circular polymers, with or without supercoiling. Quite remarkably, the torsionally relaxed and the supercoiled model DNA lead to a very similar force compression curve, which is also not far from the one we found for linear DNA. On the other hand, the dynamic response is significantly different, and markedly dependent on supercoiling. 

While the results presented in this work focus on the case of proteins which are not charged, and are the same size as the DNA beads (a choice made for simplicity), we have also performed simulations taking into account the fact that the size of a typical bacterial protein is larger, and about twice the thickness of DNA~\cite{phillips} (Appendix A), and that proteins may have a non-negligible charge (Appendix B). The size results are shown in Appendix A, Figures~\ref{pic10}, \ref{pic11} and \ref{pic12}: these correspond to Figures~\ref{pic4}, \ref{pic5} and \ref{pic6}, and were obtained with a protein size of $2\sigma$. The charge results are shown in Appendix B, Figures~\ref{pic13}, \ref{pic14}, \ref{pic15}: these correspond to Figures~\ref{pic3}, \ref{pic4}, \ref{pic5}.
In all cases, the trends are qualitatively identical, and confirm our conclusions above. There are quantitative differences for the size simulations when the volume fraction is high: this is expected as, under those situation, it is the volume fraction, rather than number density, which determines the compression pressure and force. It is interesting that the effective exponent for DNA expansion slightly decreases for larger proteins, due to the increased crowding.

Overall, our results provide a generic framework within which to analyse experiments such as those in Ref.~[\onlinecite{Pelletier}]; they also provide further testable predictions for future experiments, e.g., probing the dynamics of bacterial DNA or of supercoiled plasmids {\it in vitro}.

\begin{acknowledgments}
This work was partially funded by the European Research Council (ERC Consolidator Grant, THREEDCELLPHYSICS, Ref. 648050). MCP acknowlegdes studentship funding from EPSRC under grant no. EP/L015110/1. JL was supported by EU intra-European fellowship623637 DyCoCoS FP7-PEOPLE-2013-IEF.
\end{acknowledgments}

\appendix

\section{\label{sec:appendixA}Effect of protein size}

Here we show results corresponding to Figures~\ref{pic4}, \ref{pic5} and \ref{pic6} in the main text, but with a protein size of $2\sigma$.

\begin{figure}
\begin{center}
\includegraphics[width=8.4cm]{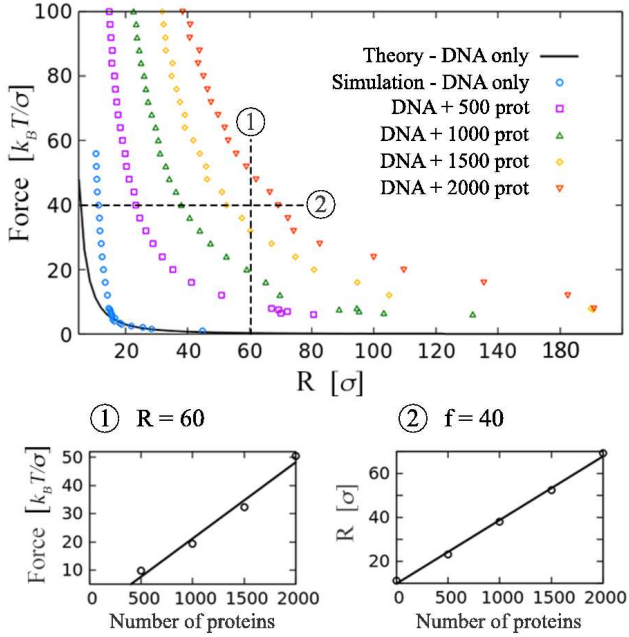}
\end{center}
\caption{\label{pic10}As Figure~\ref{pic4}, but for simulations with proteins of diameter $2\sigma$. -- Compression force as a function of the DNA extension for a varying number $M$ of non-DNA-binding proteins: DNA-only simulations (blue circles), $M=500$ (magenta squares), $M=1000$ (green triangles), $M=1500$ (yellow diamonds) and $M=2000$ (orange inverted triangles). As for the simulations presented in Figure~\ref{pic4}, the protein osmotic contribution leads to a large increase in the compression force, sometimes of several orders of magnitude. For a given value of $R$, the protein contribution is linear in the number of proteins (inset 1) -- fit: $f=0.0271M-5.8389$. For a given compression force, the DNA extension also increases linearly with the increasing number of proteins (inset 2) -- fit: $f=0.0290M+9.8640$.}
\end{figure}

Figure~\ref{pic10} shows the compression force versus extension curve: comparison between these results and those of Figure~\ref{pic4} show a very similar trend. The quantitative values are very close together for large enough $R$, and start to deviate significantly below $R\sim 60-70$, where the volume fraction is non-negligible, and excluded volume effects are larger for the larger proteins.

\begin{figure}
\begin{center}
\includegraphics[width=8.4cm]{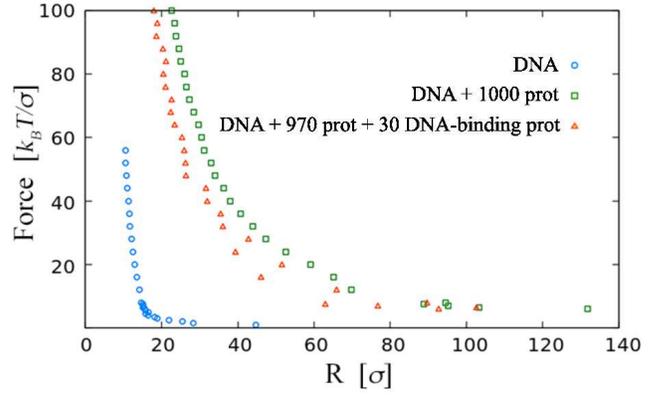}
\end{center}
\caption{\label{pic11}As Figure~\ref{pic5}, but for simulations with proteins of diameter $2\sigma$. -- Compression force as a function of the DNA extension for simulations in the absence of proteins (blue circles), in the presence of non-DNA-binding proteins (green squares) and in the presence of both binding and non-binding proteins (orange triangles: $3\%$ of the proteins are DNA-binding). Like in the simulations presented in Figure~\ref{pic5}, the DNA-binding proteins lead to the formation of DNA clusters, hence compacting the DNA, giving a decrease in the compression force. However, since here the proteins are larger, there is multiple binding to DNA, which leads to a more compact DNA, hence the slightly larger gap between the curves for the cases with and without DNA-binding proteins.}
\end{figure}

The comparison between Figure~\ref{pic11} and Figure~\ref{pic5} leads to the same conclusion -- the slightly larger gap between cases with and without DNA-binding proteins is due to the fact that larger DNA-binding proteins lead to multiple binding to the DNA, hence the DNA becomes more compact. Thus the observed effect might not just be due to protein size, as bacterial DNA-binding proteins such as H-NS only have two DNA-binding sites (they are better represented in our model by the smaller proteins considered in the main text).

\begin{figure*}
\begin{center}
\includegraphics[width=16.cm]{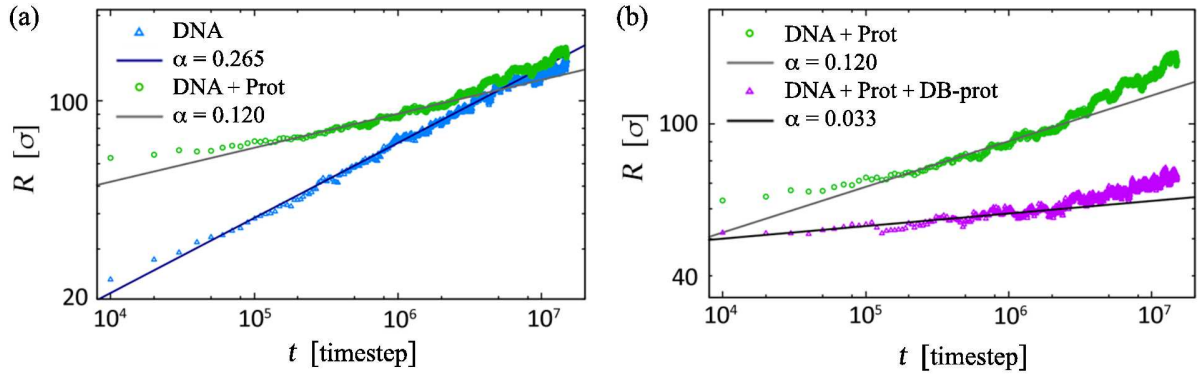}
\end{center}
\caption{\label{pic12}As Figure~\ref{pic6}, but for simulations with proteins of diameter $2\sigma$. -- Measured dynamical exponents during DNA expansion (in all cases, curves correspond to averages over 10 runs). (a) Comparison of the exponents measured for simulations without proteins (blue triangles) and with $1000$ non-DNA-binding proteins (green circles). Like in the simulations presented in Figure~\ref{pic6}, the protein crowding leads to a lower exponent, however here the exponent decrease is larger. This is due to the increased effective friction generated by larger proteins. The line for the case with proteins was fit for the range of times $[10^4:3.3\times10^6]$. (b) Comparison of the exponents measured for simulations in the presence of $1000$ non-DNA-binding proteins (green circles) and in the presence of $970$ non-DNA-binding and $30$ DNA-binding proteins (magenta triangles). The exponent decreases in the presence of DNA-binding proteins, like in the case of Figure~\ref{pic6}, however here large proteins lead to a more marked decrease of the exponent. This is due to the fact that larger proteins allow multiple binding to DNA, which further decreases the DNA extensional elasticity. The line for the case with DNA, non-DNA-binding proteins and DNA-binding proteins was fit for the range of times $[10^4:3.3\times10^6]$. All simulations started from an initial DNA configuration obtained for a compression force of $20$. The range of times, for which the expansion curves were fit, were chosen to take into account just the first expansion regime.}
\end{figure*}

The effect of the larger protein size on the expansion dynamics is analysed in Figure~\ref{pic12}, which shows a slower expansion dynamics. This is because the larger spheres create larger effective friction (as the friction, or viscosity of a hard sphere suspension is proportional to its volume fraction). The effective exponent we find is accordingly smaller, although this should be seen as a measure of the speed of the dynamics rather than a true dynamical exponent, which would require the study of different chain lengths. The decrease in the effective exponent is more marked in the case with DNA-binding proteins: again, this is due to the fact that these proteins can form multiple contacts with DNA.

\section{\label{sec:appendixB}Effect of charge}

Here we present the results corresponding to Figures~\ref{pic3}, \ref{pic4} and \ref{pic5} in the main text, but for charged DNA beads and proteins.

The electrostatic interaction between particles is modeled by considering the Debye-H\"{u}ckel potential \cite{Dwyer1995,Kunze2000}, in addition to the potentials already described in Section~\ref{sec:model},
\begin{eqnarray}
U_{{\rm DH}}(r_{i,i+1})=C\frac{q_iq_j}{\epsilon_r}\frac{e^{ka}}{1+ka}\frac{e^{-kr_{i,i+1}}}{r_{i,i+1}},
\end{eqnarray}
where $C=1/{4\pi \epsilon_0 k_BT}$, $q_i$ is the charge of particle $i$, $\epsilon_r$ the dimensionless dielectric constant (we consider $\epsilon_r=\epsilon_{r\rm{,\ water}}=80$), $k$ the inverse Debye length, and $a$ the radius of the particle.

We consider $k^{-1}=1$ nm, which is the Debye length in the cell interior. More explicitly, $k=\sqrt{8\pi l_B N_A 10^3 c_S}$, where $l_B=0.71$ is the Bjerrum length in water and $c_S\sim 150$ mM is the salt concentration inside cells \cite{Ando2010}.

Regarding the charge of the particles, DNA is known to carry two negative charges ($-2e$) per base-pair, due to the negatively charged phosphate group. As detailed in Section~\ref{sec:units}, $1$ DNA bead corresponds to $7.4$ bp in our model. Therefore, there are $7.4$ phosphate groups per DNA bead. However, counterion (or salt) condensation leads to a neutralisation of $80\%-100\%$ of the phosphate groups \cite{Forrey2006}. So, in fact, each DNA bead will only carry a charge of $0.2\times-14.8e=-2.96e$. The choice of the value of the proteins' charge is not as straight forward since it depends on the protein residues. We opted to consider the value of the charge of an average protein, and use that value for negatively and positively charged proteins in our model. In~[\onlinecite{Runcong2001}], the average of protein charges in bacteria was measured to lie in the range $[-10e,+15e]$. For simplification we consider the proteins' charge to be equal in magnitude to the DNA beads' charge -- $14.8e$ -- which corresponds to an effective charge $|q_{prot}|=2.96e$ due to neutralisation emerging from couterion condensation. Therefore, non-DNA-binding proteins are modeled as beads with charge $q=-2.96e$ and DNA-binding proteins with charge $q=+2.96e$.

\begin{figure*}
\begin{center}
\includegraphics[width=16.cm]{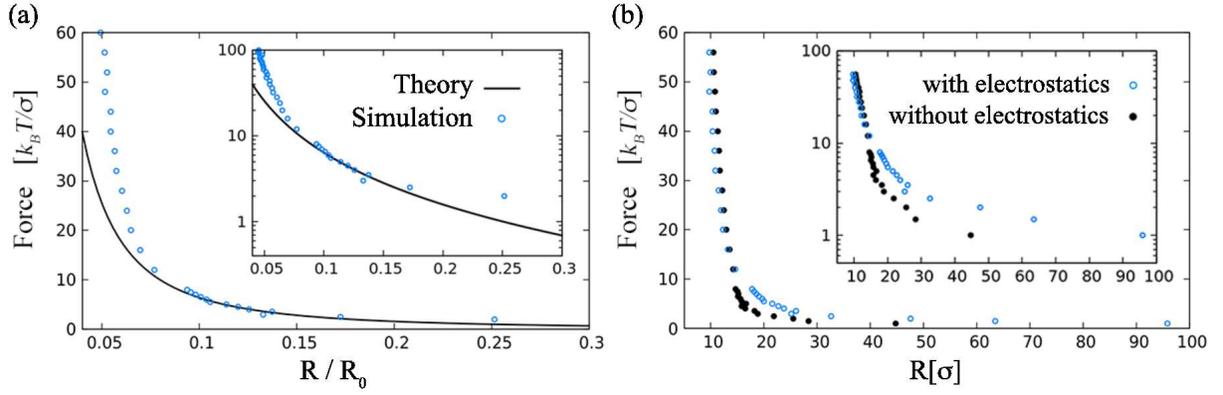}
\end{center}
\caption{\label{pic13}As Figure~\ref{pic3}, but for simulations with charged proteins and polymer beads. -- Compression force as a function of the normalized DNA extension in the absence of proteins. (a) Comparison of the force-extension curve obtained from the entropic spring theory of Ref.~[\onlinecite{Pelletier}] (full line) and our numerical model (blue circles). Like in the simulations presented in Figure~\ref{pic3}, the theory of an infinitesimally thin polymer agrees with the model for $R/R_0>0.1$, but breaks down for $R/R_0<0.1$, where excluded volume effects become significant. The inset shows the results for the ``Force'' axis in logarithmic scale. (b) Comparison of the force-extension curves obtained from the model with (blue circles) and without (black dots) electrostatic interactions. The inset shows the force axis in logarithmic scale. For $f>10$, both models lead to similiar force-extension relations (for very large forces, the case with electrostatic interactions is slightly smaller. However this only holds for a single realisation). For $f<10$, the extension of the polymer in the presence of electrostatic interactions for a given force is larger than for the case without electrostatics, which was expected since there is an additional long-range repulsion between DNA beads.}
\end{figure*}

Figure~\ref{pic13} shows the compression force versus extension in the absence of proteins. Figure~\ref{pic13}(a) shows that, like in Figure~\ref{pic3}, the entropic spring theory of Ref.~[\onlinecite{Pelletier}] agrees with the model for $R/R_0>0.1$, but breaks down for $R/R_0<0.1$. Figure~\ref{pic13}(b) shows that, for $f>10$, the DNA beads' charge does not play a significant role in the elastic response of the DNA polymer, but that for weak compression forces ($f<10$) the effect of long-range electrostatic interactions is more noticeable: for a given force the DNA extension for the charged polymer is larger than for the neutral polymer.

\begin{figure}
\begin{center}
\includegraphics[width=8.4cm]{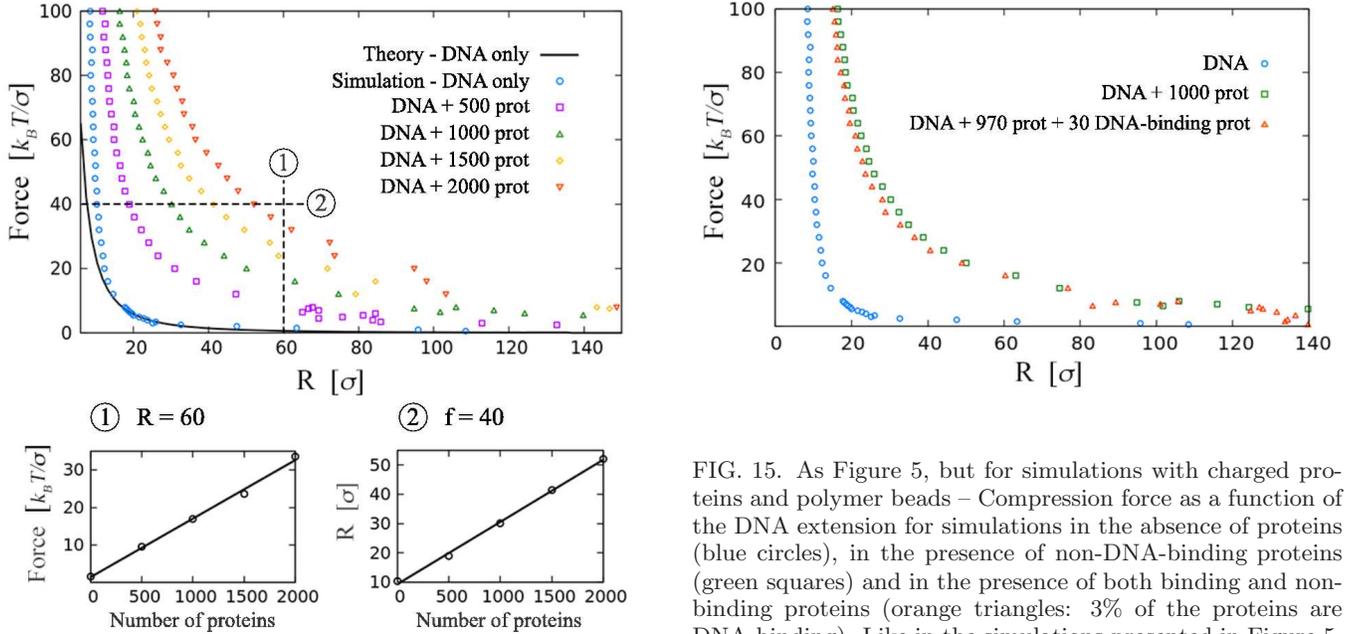}
\end{center}
\caption{\label{pic14}As Figure~\ref{pic4}, but for simulations with charged proteins and polymer beads -- Compression force as a function of the DNA extension for a varying number $M$ of non-DNA-binding proteins: DNA-only simulations (blue circles), $M=500$ (magenta squares), $M=1000$ (green triangles), $M=1500$ (yellow diamonds) and $M=2000$ (orange inverted triangles). The force-extension curves are remarkably similar to the ones in Figure~\ref{pic4}. The linear relations force - $M$ (inset 1 -- fit: $f=0.0156M+1.4513$), and DNA extension - $M$ (inset 2 -- fit: $f=0.0212M-9.3741$) are also recovered. The comparison between inset 2 and the one in Figure~\ref{pic4} shows that the presence of electrostatic interactions leads to a slightly higher DNA extension for a moderate force, as seen in Figure~\ref{pic13}.}
\end{figure}

Figure~\ref{pic14} for charged particles and Figure~\ref{pic4} for neutral particles are remarkably similar. The comparison between inset 2 in both Figures shows, however, that the presence of electrostatic interactions leads to a slightly higher DNA extension for a moderate force, as seen before.

\begin{figure}
\begin{center}
\includegraphics[width=8.4cm]{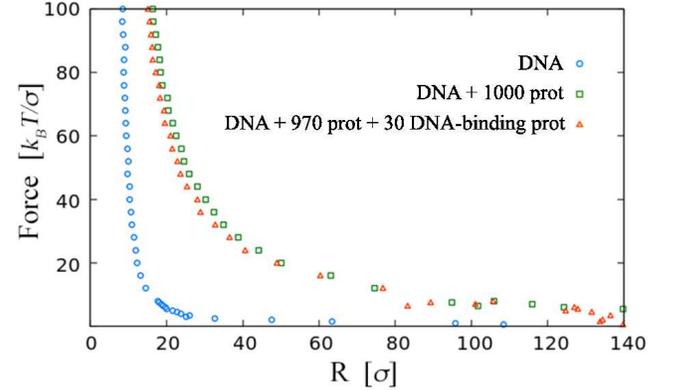}
\end{center}
\caption{\label{pic15}As Figure~\ref{pic5}, but for simulations with charged proteins and polymer beads -- Compression force as a function of the DNA extension for simulations in the absence of proteins (blue circles), in the presence of non-DNA-binding proteins (green squares) and in the presence of both binding and non-binding proteins (orange triangles: $3\%$ of the proteins are DNA-binding). Like in the simulations presented in Figure~\ref{pic5}, the DNA-binding proteins lead to the formation of DNA clusters, hence compacting the DNA, giving a decrease in the compression force. Again we point out the striking similarity between these curves and the ones in Figure~\ref{pic5}. This suggests that, indeed, the electrostatic interactions don't play a significant role in the overall elastic response of DNA.}
\end{figure}

Again, Figure~\ref{pic15} and Figure~\ref{pic5} lead to the same conclusions. In both cases the DNA-binding proteins lead to the formation of DNA clusters, hence compacting the DNA, giving a decrease in the compression force. The striking similarity between the results arising from the charged and neutral models further suggests that the electrostatic interactions do not play a significant role in the overall elastic response of DNA.

\section*{References}
\nocite{*}
\bibliography{bibliography}

\end{document}